\documentclass[aps,pra,twocolumn,showpacs,floatfix]{revtex4}
\usepackage{epsfig}
\usepackage{graphicx}
\usepackage{dcolumn}
\usepackage{amsthm,amsmath}

\begin{document}

\title{Relativistic Equation of Motion Coupled-Cluster Method: Application to the closed-shell atomic systems} 
\author{Himadri Pathak\footnote{h.pathak@ncl.res.in}$^1$, B. K. Sahoo$^2$, B. P. Das$^3$, Nayana Vaval$^1$, Sourav Pal\footnote{s.pal@ncl.res.in}$^1$, }
\affiliation{$^1$Electronic Structure Theory Group, Physical Chemistry Division, CSIR-National Chemical Laboratory, Pune, 411008, India}
\affiliation{$^2$Theoretical Physics Division, Physical Research Laboratory, Ahmedabad, 380009, India}
\affiliation{$^3$Theoretical Physics and Astrophysics Group, Indian Institute of Astrophysics, Bangalore-560034, India}

\begin{abstract}

We report our successful implementation of the full fledged relativistic equation of motion coupled cluster (EOMCC) 
method. This method is employed to compute the principal ionization potentials (IPs) of closed-shell rare 
gas atoms, He-like ions, Be-like ions along with Na$^+$, Al$^+$, K$^+$, Be, and Mg. Four component Dirac spinors are
used in the calculations and the one and two electron integrals are evaluated using the Dirac Coulomb Hamiltonian.
Our results are in excellent agreement with those available measurements, which are taken from the National Institute of Science and Technology database (NIST).
We also present results using the second order many-body perturbation theory (MBPT(2)) and random phase approximation (RPA)
in the EOMCC framework. These results are compared with those of EOMCC at the level of single and double excitations in order
to assess the role of the electron correlation effects in the intermediate schemes considered in our calculations .
\end{abstract}
\maketitle
\section{Introduction}
  High precision calculations of the spectroscopic properties of heavy atomic and molecular systems are challenging due to the complex 
interplay between relativistic and correlation effects \cite{0}. However, with the extension of several well established 
non relativistic many-body methods to the relativistic regime and the recent advances in high performance computing techniques, such 
calculations are no longer insurmountable. Studies of atomic parity non-conservation (PNC) and permanent electric 
dipole moments (EDMs) due to the violation of parity and time reversal symmetries \cite{1,2}, requirement of very accurate 
atomic properties for the precise estimate of systematic effects in the atomic clock experiments \cite{3,4,5}, determination of 
nuclear moments \cite{6}, calculations of sensitive coefficients to probe the variation of the fine structure 
constant \cite{7,7a,8} etc. require the development of powerful relativistic many-body methods. The spectra of multi-charged ions are 
of immense interest in many areas of physics; particularly in x-ray space astronomy, plasma physics and laser
physics \cite{9,10}. Accurate values of ionization potentials (IPs), double ionization potentials 
(DIPs), and excitation energies (EEs), especially from the deep core orbitals, are required for setting up the probe and 
its tunability of the ionizing beam in experiments like e-2e, e-3e, $\gamma$-2e, double Auger decay etc \cite{10a,10b}.

  Among the various wave function based methods, the coupled-cluster (CC) theory within the singles and
doubles (CCSD) approximation is the most elegant way of calculating energy or
energy differences of atoms and molecules in the ground state as well as in the excited states \cite{11}.
Green's function and propagator techniques \cite{12,13} are the two
traditional approaches to calculate direct energy differences. In the propagator approaches, the ground and excited 
states are treated simultaneously and due to the cancellation of common correlation effects, these approaches provide
satisfactory results of these energy differences in a direct manner. In the CC domain, the Fock space multi-reference 
CC (FSMRCC) \cite{14,15,16,17,18,19,20} and the EOMCC method \cite{21,22,23} are 
the two most familiar variants for the calculation of direct energy differences. Many non-relativistic calculations of
IPs, and EEs both in the FSMRCC \cite{26} and EOMCC \cite{28,28a,28b} frameworks are available, but their relativistic 
counterparts are far fewer for the former case and none for the latter.

\begin{table*}[t]
\caption{The $\mathrm{\alpha_{0}}$ and $\mathrm{\beta}$ parameters of the even tempered basis used in calculations.} 
\begin{ruledtabular}
\begin{tabular}{lcccccccccccccccccccc}
     Atom &  \multicolumn{2}{c}{$\mathrm{s}$}  & \multicolumn{2}{c}{$\mathrm{p}$}  &
     \multicolumn{2}{c}{$\mathrm{d}$}  & \multicolumn{2}{c}{$\mathrm{f}$} & \multicolumn{2}{c}{$\mathrm{g}$}\\
     & $\mathrm{\alpha_{0}}$  & $\mathrm{\beta}$ & $\mathrm{\alpha_{0}}$ & $\mathrm{\beta}$  
     & $\mathrm{\alpha_{0}}$  & $\mathrm{\beta}$ & $\mathrm{\alpha_{0}}$& $\mathrm{\beta}$ &$\mathrm{\alpha_{0}}$&$\mathrm{\beta}$\\
\cline{2-3} \cline{4-5} \cline{6-7} \cline{8-9} \cline{10-11} \\
    \hline
He  &\,0.00075  &\,2.075  &\,0.00155&\,2.080  &\,0.00258 &\,2.180  &\,0.00560 &\,2.300  &\,0.00765 &\,2.450  \\
Li  &\,0.00750  &\,2.075  &\,0.00755&\,2.070  &\,0.00758 &\,2.580  &\,0.00760 &\,2.600  &\,0.00765 &\,2.650  \\
Be  &\,0.00500  &\,2.500  &\,0.00615&\,2.650  &\,0.00505 &\,2.550  &\,0.00500 &\,2.530  &\,0.00480 &\,2.500  \\
Ne  &\,0.00753  &\,2.075  &\,0.00755&\,2.070  &\,0.00758 &\,2.580  &\,0.00800 &\,2.720  &\,0.00800 &\,2.720  \\
Na  &\,0.00250  &\,2.210  &\,0.00955&\,2.215  &\,0.00700 &\,2.750  &\,0.00710 &\,2.760  &\,0.00715 &\,2.765  \\
Mg  &\,0.02950  &\,1.630  &\,0.09750&\,1.815  &\,0.00750 &\,2.710  &\,0.00780 &\,2.730  &\,0.00800 &\,2.750  \\  
Ar  &\,0.09850  &\,1.890  &\,0.00720&\,2.965  &\,0.00700 &\,2.700  &\,0.00700 &\,2.690  &\,0.00700 &\,2.696  \\
K   &\,0.00550  &\,2.250  &\,0.00995&\,2.155  &\,0.00690 &\,2.550  &\,0.00700 &\,2.600  &\,0.00700 &\,2.600  \\                       
Kr  &\,0.00020  &\,2.022  &\,0.00720&\,2.365  &\,0.00700 &\,2.550  &\,0.00700 &\,2.695  &\,0.00700 &\,2.695  \\
Xe  &\,0.00010  &\,2.022  &\,0.00720&\,2.365  &\,0.00700 &\,2.550  &\,0.00700 &\,2.695  &\,0.00700 &\,2.695  \\ 
Rn  &\,0.00010  &\,2.280  &\,0.00671&\,2.980  &\,0.00715 &\,2.720  &\,0.00720 &\,2.710  &\,0.00720 &\,2.695  \\
\end{tabular}
\end{ruledtabular}
\label{tabgto}
\end{table*}

 Relativistic calculations are necessary for the spectral properties of heavy atoms and molecules as well as for highly stripped
heavy ions. It 
 is therefore desirable in such cases to have a theory which can simultaneously 
treat the electron correlation and the effects of relativity on the same footing as they are non-additive in nature. Kaldor and coworkers 
were the first to develop a relativistic coupled-cluster theory for this purpose. They applied the  relativistic FSMRCC method to atoms 
as well as molecules. \cite{31,32,33,34}. 
The effective Hamiltonian formalism of the FSMRCC theory, based on the Bloch equation, acts within a model space 
\cite{35,36}. It uses a common vacuum with respect to which holes and particles are defined. The holes and particles are 
further classified into active and inactive depending on the requirements of the problem. While an increase in the size of the 
model space can target more states, it can lead to convergence problems, 
which is well known in the literature as the intruder state problem \cite{37,38}. The EOMCC method is basically single reference in
nature and is closely related to the CC linear response theory (CCLRT) \cite{39,40,41}. Chaudhuri et al. had
applied the relativistic CCLRT to the ionization problem \cite{42}. Hirata and co-workers \cite{44} had employed the
relativistic EOMCC method using two component valence pseudo-spinors alongwith a relativistic effective core 
potential (RECP) which was supplemented by the spin-orbit interaction \cite{45}. The approach of Hirata and coworkers clearly
lacks a rigorous 
description of the relativistic effects, which can be taken into account by using four component single particle wave functions and the
Dirac Coulomb Hamiltonian.

  In the present work, we consider the EOMCC method in the four component relativistic framework within the singles and
doubles approximation (EOM-CCSD method) to calculate IPs by removing one electron from a closed-shell 
atomic system. This EOMCC method for the ionization
problem is size consistent, and is equivalent to the (0,1) sector of the FSMRCC theory \cite{46,47}. It is capable of
providing the principal as well as shake up IP values. The (0,1) sector FSMRCC theory does not address the
shake-up states. Though the EOM-CCSD method is a size extensive method for the principal valence sector 
\cite{48,49}, it is not so for the shake-up states. The error due to the size extensivity is reduced due to the presence of the
two hole-one particle (2h-1p) block. Being an eigenvalue problem, it is not affected by numerical instabilities due to the intruder states, which are
very common in the FSMRCC method, do not arise.  
Two intermediate calculations are employed to assess the effects of electron correlation. We refer these as
EOM-MBPT(2) and EOM-RPA methods. The scheme named EOM-MBPT(2)
uses a first order perturbed ground state wave function which corresponds
to MBPT(2) energy as ground state energy and in the scheme EOM-RPA, the EOM matrix elements are
constructed in the one-hole (1h-0p) space.

This paper is  organized as follows. A brief discussion of the relativistic method used to obtain the 
single particle orbitals is presented in Sec. \ref{sec2}. This is followed by a description of the EOMCC theory of the
ionization problem and the computational details are presented in Sec. \ref{sec3}. In Sec. \ref{sec4}, we give our results 
and discuss them before making our concluding remarks in Sec \ref{sec5}. Unless stated otherwise we have used
atomic units (a.u.) through out the paper.

\begin{table}[t]
\caption{SCF energy ($\mathrm{E_{DF}^{0}}$), correlation energies from the MBPT(2) ($\mathrm{E_{corr}^{(2)}}$) and
CCSD ($\mathrm{E_{corr}^{(ccsd)}}$) methods along with the numbers of active orbitals from various 
symmetries taken in the calculations for different atoms.}
\begin{ruledtabular}
%\begin{tabular}{lccccrrrccccr}
\begin{tabular}{lrrrrrrrrrrrr}
Atom & \multicolumn{5}{c}{No. of active orbitals} & {$\mathrm{E_{DF}^{0}}$} & 
{$\mathrm{E_{corr}^{(2)}}$}& {$\mathrm{E_{corr}^{(ccsd)}}$}  \\
 \cline{2-6} \\
& $\mathrm{s}$  & $\mathrm{p}$ & $\mathrm{d}$ & $\mathrm{f}$  
& $\mathrm{g}$   \\
\hline
& & \\
$\mathrm{He}$        &\,16   &\, 14   &\, 12  &\, \, 9  &\, 7  &\,   -2.8618  &\,  -0.0365  &\, -0.0415\\
$\mathrm{Li^{+}}$    &\,15   &\, 14   &\, 10  &\, \, 9  &\, 8  &\,   -7.2372  &\,  -0.0395  &\, -0.0430\\
$\mathrm{Ne^{8+}}$   &\,16   &\, 15   &\, 11  &\,\, \,9  &\, 8  &\,  -93.9827  &\,  -0.0421  &\, -0.0434\\
$\mathrm{Na^{9+}}$   &\,16   &\, 15   &\, 13  &\, 10  &\, 9  &\, -114.4158  &\,  -0.0414  &\, -0.0426\\
$\mathrm{Ar^{16+}}$  &\,14   &\, 11   &\, 11  &\, 10  &\, 8  &\, -314.1995  &\,  -0.0409  &\, -0.0417\\
$\mathrm{Kr^{34+}}$  &\,22   &\, 13   &\, 11  &\, 10  &\, 9  &\,-1296.1641  &\,  -0.0237  &\, -0.0240\\
$\mathrm{Be}$        &\,13   &\, 11   &\, 11  &\, \, 9  &\, 8  &\,  -14.5758  &\,  -0.0742  &\, -0.0924\\
$\mathrm{B^{+}}$     &\,15   &\, 14   &\, 10  &\, \, 9  &\, 8  &\,  -24.2451  &\,  -0.0824  &\, -0.1062\\                       
$\mathrm{C^{2+}}$    &\,15   &\, 13   &\, 11  &\, 10  &\, 9  &\,  -36.4251  &\,  -0.0924  &\, -0.1215\\
$\mathrm{N^{3+}}$    &\,15   &\, 14   &\, 13  &\, 10  &\, 9  &\,  -51.1144  &\,  -0.1026  &\, -0.1369\\ 
$\mathrm{O^{4+}}$    &\,15   &\, 14   &\, 12  &\, 10  &\, 9  &\,  -68.3143  &\,  -0.1089  &\, -0.1487\\
$\mathrm{F^{5+}}$    &\,15   &\, 14   &\, 13  &\, 10  &\, 9  &\,  -88.0271  &\,  -0.1168  &\, -0.1621\\
$\mathrm{Ne^{6+}}$   &\,16   &\, 15   &\, 13  &\, 10  &\, 9  &\, -110.2559  &\,  -0.1237  &\, -0.1744\\
$\mathrm{Na^{7+}}$   &\,15   &\, 14   &\, 11  &\, 10  &\, 9  &\, -135.0042  &\,  -0.1266  &\, -0.1829\\
$\mathrm{Mg^{8+}}$   &\,15   &\, 14   &\, 13  &\, 11  &\, 9  &\, -162.2763  &\,  -0.1352  &\, -0.1966\\
$\mathrm{Al^{9+}}$   &\,15   &\, 14   &\, 13  &\, 10  &\, 9  &\, -192.0767  &\,  -0.1404  &\, -0.2072\\
$\mathrm{Si^{10+}}$  &\,15   &\, 14   &\, 13  &\, 11  &\, 9  &\, -224.4105  &\,  -0.1461  &\, -0.2177\\
$\mathrm{P^{11+}}$   &\,15   &\, 14   &\, 13  &\, 11  &\,10  &\, -259.2833  &\,  -0.1513  &\, -0.2278\\
$\mathrm{S^{12+}}$   &\,15   &\, 14   &\, 13  &\, 11  &\,10  &\, -296.7011  &\,  -0.1561  &\, -0.2374\\
$\mathrm{Cl^{13+}}$  &\,15   &\, 14   &\, 13  &\, 11  &\, 9  &\, -336.6703  &\,  -0.1606  &\, -0.2466\\
$\mathrm{Ar^{14+}}$  &\,15   &\, 14   &\, 13  &\, 11  &\,10  &\, -379.1979  &\,  -0.1650  &\, -0.2554\\
$\mathrm{Kr^{32+}}$  &\,16   &\, 15   &\, 14  &\, 11  &\,10  &\,-1593.0492  &\,  -0.2316  &\, -0.3630\\
$\mathrm{Ne}$        &\,17   &\, 17   &\, 13  &\, 11  &\,10  &\, -128.6919  &\,  -0.3736  &\, -0.3732\\
$\mathrm{Na^{+}}$    &\,17   &\, 15   &\, 11  &\, 10  &\, 9  &\, -161.8958  &\,  -0.3691  &\, -0.3715\\
$\mathrm{Mg}$        &\,20   &\, 14   &\, 11  &\, 10  &\, 8  &\, -199.9350  &\,  -0.4074  &\, -0.4174\\
$\mathrm{Al^{+}}$    &\,15   &\, 14   &\, 13  &\, 10  &\, 9  &\, -242.1290  &\,  -0.3951  &\, -0.4065\\
$\mathrm{Ar}$        &\,14   &\, 11   &\, 11  &\, 10  &\, 8  &\, -528.6657  &\,  -0.6513  &\, -0.6640\\
$\mathrm{K^{+}}$     &\,15   &\, 14   &\, 12  &\, 10  &\, 8  &\, -601.3780  &\,  -0.6664  &\, -0.6799\\
$\mathrm{Kr}$        &\,22   &\, 13   &\, 11  &\, \, 9  &\, 8  &\,-2788.8492  &\,  -1.5247  &\, -1.4622\\
$\mathrm{Xe}$        &\,23   &\, 13   &\, 12  &\, \, 9  &\, 7  &\,-7446.8108  &\,  -2.1180  &\, -2.0009\\
$\mathrm{Rn}$        &\,21   &\, 13   &\, 12  &\, 10  &\, 9  &\-23595.8070  &\,  -3.7880  &\, -3.4583\\
\end{tabular}
\end{ruledtabular}
\label{tabcorr}
\end{table}

\section{Generation of relativistic orbitals}\label{sec2}

The four component DC Hamiltonian used to evaluate the atomic integrals is given by 
\begin{eqnarray}
H &=& \sum_i \left [ c\mbox{\boldmath$\alpha$}_i\cdot \textbf{p}_i+(\beta_i -1)c^2 + V_{nuc}(r_i) +
\sum_{j>i} \frac{1}{r_{ij}} \right ], \ \ \
\end{eqnarray}
where $\mbox{\boldmath$\alpha$}_i$ and $\beta_i$ are the usual Dirac matrices, $V_{nuc}(r_i)$ is the nuclear 
potential and $\frac{1}{r_{ij}}=\frac{1}{\mathrm{\vec r}_i - \mathrm{\vec r}_j}$ is the electron-electron repulsion potential. 
Subtraction of the identity operator from $\beta$ means 
that the energies are scaled with reference to the rest mass energy of the electron. The nuclear potential is evaluated 
considering the Fermi-charge distribution of the nuclear density as given by
\begin{equation}
\rho_{nuc}(r)=\frac{\rho_{0}}{1+e^{(r-c)/a}}
\end{equation}
where the parameter '$c$' is the half-charge radius as $\rho_{nuc}(r)=\rho_0/2$ for $r=c$ and '$a$' is related 
to the skin thickness which are evaluated by
\begin{eqnarray}
a&=& 2.3/4(ln3) \\
\text{and} \ \ \ \ \ c&=& \sqrt{\frac {5}{3} r_{rms}^2 - \frac {7}{3} a^2 \pi^2}
\end{eqnarray}
with $r_{rms}$ is the root mean square radius of the nucleus.

 In relativistic quantum mechanics, the four component single particle electron orbital is given by
\begin{eqnarray}
 |\phi(r) \rangle = \frac {1}{r} \left ( \begin{matrix} P(r) & \chi_{\kappa,m}(\theta,\phi)  \cr
                                                  iQ(r) & \chi_{-\kappa,m}(\theta,\phi) \cr
                                                        \end{matrix}  \right )
\end{eqnarray}
where $P(r)$ and $Q(r)$ are the large and small components of the wave function and the angular functions are given by
\begin{eqnarray}
\chi_{\kappa,m}(\theta, \phi) = \sum_{\sigma=\pm \frac {1}{2}} C( l \sigma j; m-\sigma, \sigma) Y_l^{m-\sigma}(\theta, \phi) \phi_{\sigma}
\end{eqnarray}
for the Clebsch-Gordan (Racah) coefficient $C(l \sigma j; m-\sigma, \sigma )$, the normalized spherical harmonics 
$Y_l^{m-\sigma}(\theta, \phi)$, the Pauli two-component spinors $\phi_{\sigma}$ and the relativistic quantum number
$\kappa = -(j+\frac {1}{2})a$ satisfying the condition for the orbital angular momentum $l = j - \frac{a}{2}$ with 
the total angular momentum $j$.

\begin{table}[t]
\caption{Convergence pattern of ionization potentials of
Be atom (in ev) as a function of the active orbitals using EOM-CCSD method.} 
\begin{ruledtabular}
\begin{tabular}{lcc}
No of active orbitals  & \multicolumn{2}{c}{IP values}\\
\cline{2-3}\\
                & $\mathrm{1s}$ & $\mathrm{2s}$\\
\hline 
& & \\
91 (13s,11p,11d,9f,8g)&\,124.6463&\,9.3247\\
100 (14s,12p,12d,10f,9g)&\,124.6565&\,9.3248\\
109 (15s,13p,13d,11f,10g)&\,124.6620&\,9.3249\\
116 (16s,14p,13d,12f,11g)&\,124.6639&\,9.3249\\
118 (16s,14p,14d,12f,11g)&\,124.6639&\,9.3248\\   
\end{tabular}
\end{ruledtabular}
\label{tabbe}
\end {table}

To generate the single particle orbitals, we use the relativistic Hartree-Fock (Dirac-Fock (DF)) Hamiltonian 
given by
\begin{eqnarray}
H_{DF} &=&\sum_{j} [c \ \vec \alpha \cdot \vec p_{j} + (\beta -1)c^{2} + V_{nuc}(r_j) + U(r_j)] \nonumber \\
       &=& \sum_j h_0(r_j)
\end{eqnarray}
where $h_0$ is the single particle Fock operator with the DF potential
\begin{eqnarray}
U|\phi_j\rangle = \sum_{a=1}^{occ} \langle \phi_a |\frac{1}{r_{ja}}|\phi_a \rangle |\phi_j \rangle
 - \langle \phi_a |\frac{1}{r_{aj}}|\phi_j \rangle |\phi_a \rangle
\end{eqnarray}
for all the occupied orbitals $occ$ that leaves out contributions from the residual interaction $V_{es}= \sum_{j<l}  \frac{1}{r_{jl}} - \sum_{j} U(r_j)$ which 
is incorporated through the EOMCC method. 

 To retain the atomic spherical symmetry property in our calculations, the matrix form of the Coulomb interaction operator using the 
above single particle wave functions are expressed as
\begin{eqnarray}
\langle \phi_a \phi_b  | \frac {1}{r_{12}} | \phi_c \phi_d \rangle &=& \int dr_1 [P_a(r_1)P_c(r_1) + Q_a(r_1)Q_c(r_1)] 
 \nonumber \\ & \times & \int dr_2 [P_b(r_2)P_d(r_2) + Q_b(r_2)Q_d(r_2)] \nonumber \\
  & \times & \frac {r_<^k}{r_>^{k+1}}  \times Ang,  
\end{eqnarray}
with the multipole $k$ determined by $|j_a - j_c| \le k \le j_a + j_c$ and $|j_b - j_d| \le k \le j_b + j_d$. The 
angular momentum factor of the above expression is given by 
\begin{eqnarray}
Ang &=& \delta(m_a-m_c,m_d-m_b) \sum_k \Pi^e(\kappa_a,\kappa_c,k)  \nonumber \\ & \times & \Pi^e(\kappa_b,\kappa_d,k) 
d^k(j_cm_c,j_am_a) d^k(j_bm_b,j_dm_d), \ \ \ \
\end{eqnarray}
where the coefficient $d^k(jm,j'm')$ is defined as
\begin{eqnarray}
d^k(jm,j'm') &=& (-1)^{m+\frac {1}{2}} \frac {[(2j+1)(2j'+1)]^{\frac {1}{2}}}{(2k+1)} \nonumber \\
 && \times C(jkj';\frac {1}{2},-\frac {1}{2})C(jkj';-m,m') \ \ \
\end{eqnarray}
with $\Pi^e(\kappa,\kappa',k) = \frac {1}{2} [1-aa'(-1)^{j+j'+k}]$ for $l + l' + k =$ even.

The DF single particle orbitals, $|\phi_{n,\kappa} (r) \rangle$s, with principal quantum number $n$ and angular 
quantum number $\kappa$ are initially constructed as linear combinations of Gaussian type of orbitals (GTOs) by
writing
\begin{eqnarray}
 |\phi_{n,\kappa} (r) \rangle = \frac {1}{r} \sum_{\nu} \left (
         \begin{matrix}
         C_{n,\kappa}^L N_L f_{\nu}(r) & \chi_{\kappa,m} \cr
         i C_{n, -\kappa}^S N_S \left (\frac{1}{dr} + \frac{\kappa}{r} \right ) f_{\nu}(r) &\chi_{-\kappa,m}\cr
                 \end{matrix}
         \right ), \ \
\end{eqnarray}
where $C_{n,\kappa}$s are the expansion coefficients, $N_{L(S)}$ is the normalization constant for the large
(small) component of the wave function and $\alpha_{\nu}$ is a suitably chosen parameter for orbitals of different 
angular momentum symmetries and $f_{\nu}(r) = r^l e^{-\alpha_{\nu} r^2}$ is a GTO. For the exponents, we use the even 
tempering condition $\alpha_{\nu} = \alpha_0 \beta^{\nu-1}$ with two parameters $\alpha_0$ and $\beta$. It  can be 
noticed in the above expression that the large and small components of the wave function satisfy the kinetic balance 
condition. The orbitals are finally obtained after solving the matrix eigenvalue form of the DF equation by a self-
consistent procedure.

\begin{table}[t]
\caption{Ionization potentials (IPs) of helium (He) like systems in eV using MBPT(2), RPA and CCSD methods in the EOM procedure.}
\begin{ruledtabular}
%\begin{tabular}{lcccc}
\begin{tabular}{rrrrr}

Atom  & MBPT(2)  &  RPA &  CCSD  &  NIST \cite{51} \\
\hline
$\mathrm{Li^{+}}$  &\,    75.5517 &\,   77.1594  &\,   75.6399 &\,   75.6400  \\
$\mathrm{Ne^{8+}}$ &\,  1196.1770 &\, 1197.7308  &\, 1196.2113 &\, 1195.8078 \\
$\mathrm{Na^{9+}}$ &\,  1465.6073 &\, 1467.1611  &\, 1465.6401 &\, 1465.1344 \\
$\mathrm{Ar^{16+}}$&\,  4123.5442 &\, 4125.1003  &\, 4123.5661 &\, 4120.6654  \\                       
$\mathrm{Kr^{34+}}$&\, 17323.3995 &\,17324.9869  &\,17323.4104 &\, 17296.4200 \\
\end{tabular}
\end{ruledtabular}
\label{tabhe}
\end{table}

\begin{table*}[t]
\caption{Ionization potentials (IPs) of beryllium (Be) like systems in eV using MBPT(2), RPA and CCSD methods in the EOM procedure.}
\begin{ruledtabular}
%\begin{tabular}{lccccccccccc}
\begin{tabular}{rrrrrrrrrrrr}

Ion & \multicolumn{2}{c}{$\mathrm{MBPT(2)}$} & & \multicolumn{2}{c}{$\mathrm{RPA}$} & & 
     \multicolumn{2}{c}{$\mathrm{CCSD}$}& & \multicolumn{2}{c}{$\mathrm{NIST}$ \cite{51}} \\
\cline{2-3} \cline{5-6} \cline{8-9} \cline{11-12} \\
     & $\mathrm{1s}$  & $\mathrm{2s}$ & & $\mathrm{1s}$ & $\mathrm{2s}$  
     & & $\mathrm{1s}$  & $\mathrm{2s}$ & & $\mathrm{1s}$ & $\mathrm{2s}$ \\
     \hline
     & & \\
$\mathrm{B^{+}}$  &\,218.7753  &\,24.6024  & &\,223.7170 &\,25.4690 & &\, 218.6932 &\, 25.1510  & & 217.8827 &\,25.1548 \\
$\mathrm{C^{2+}}$ &\,340.5912  &\,47.1763  & &\,345.3340  &\,48.1961 & &\, 340.5074 &\, 47.8838  & & &\,47.8877 \\
$\mathrm{N^{3+}}$  &\,489.5193   &\,76.6082 & &\,494.3701  &\,77.7833 & &\, 489.3987 &\, 77.4732  & & &\,77.4735 \\
$\mathrm{O^{4+}}$ &\,665.8043  &\,112.8779  & &\,670.6873  &\,114.2098 & &\, 665.6751 &\, 113.9003 & & &\,113.8990    \\
$\mathrm{F^{5+}}$  &\,869.6607  &\,155.9937 & &\,874.4161 &\,157.4809 & &\, 869.5295 &\,157.1714 & & &\,157.1631     \\
$\mathrm{Ne^{6+}}$  &\,1100.7242  &\,205.9558 & &\,1105.5077 &\,207.5972 & &\, 1100.5835 &\, 207.2874 & & 1098.7791 &\,207.2710 \\
$\mathrm{Na^{7+}}$ &\,1359.1193  &\,262.7653 & &\,1363.9246 &\,264.5608 & &\, 1358.9780 &\,264.2504 & & 1357.1716 &\,264.1920    \\
$\mathrm{Mg^{8+}}$ &\,1644.9936  &\,326.4618 & &\,1649.9248 &\,328.4010 & &\, 1644.8387 &\, 328.0902 & & &\,327.9900   \\
$\mathrm{Al^{9+}}$ &\,1958.6549  &\,397.0176 & &\,1963.3552 &\,399.1102 & &\, 1958.5119 &\, 398.7986 & & 1955.7950 &\,398.6500  \\
$\mathrm{Si^{10+}}$ &\,2299.5858   &\,474.4895 & &\,2304.3242 &\,476.7141 & &\, 2299.4367 &\, 476.4017 & & 2296.5894 &\,476.1800  \\
$\mathrm{P^{11+}}$ &\,2668.1363  &\,558.8627 & &\,2672.8963 &\,561.2228 & &\, 2667.9846 &\, 560.9095 & & 2664.7632 &\,560.6200  \\
$\mathrm{S^{12+}}$  &\,3064.3424   &\,650.1586 & &\,3069.1229 &\,652.6532 & &\, 3064.1883 &\, 652.3391 & & 3059.9469 &\,651.9600  \\
$\mathrm{Cl^{13+}}$  &\,3488.2444  &\,748.3994 & &\,3493.0728  &\,751.0246 & &\, 3488.0867 &\, 750.7090  & & &\,750.2300   \\
$\mathrm{Ar^{14+}}$  &\,3941.3783   &\,853.6104 & &\,3944.8161  &\,856.3589  & &\, 3941.4781 &\, 856.0432 & & 3934.7226 &\,855.4700    \\
$\mathrm{Kr^{32+}}$  &\,16934.9486  &\,3972.1671 & &\,16939.9718 &\,3976.0698  & &\,16934.8134 &\,3975.7297 & & 16902.8643 &\,3971.0000  \\
\end{tabular}
\end{ruledtabular}
\label{belk}
\end{table*}

\section{Method of calculations: EOM-CCSD }\label{sec3}

In the CC method, the ground state wave function of a closed-shell atomic system is defined as 
\begin{equation}
|\Psi_{0}\rangle=e^{T}|\Phi_{0} \rangle,
\end{equation}
where $|\Phi_{0} \rangle$ is the DF wave function. 
The excited states is defined as 
\begin{eqnarray}
H |\Psi_{\mu}\rangle &=& E_{\mu}|\Psi_{\mu} \rangle =E_{\mu} R_{\mu}|\Psi_{0} \rangle,
\end{eqnarray}
for a linear CI like excitation operator $R_\mu$.  

The operators $R_{\mu}$ commute with $T$  as they are strings of quasi-particle 
creation operators (but not necessarily particle conserving). Pre-multiplying the above equation
with the non-singular operator $e^{-T}$ leads to   
\begin{equation}
[\overline{H},R_{\mu}]|\Phi_{0} \rangle= \Delta E_{\mu}R_{\mu}|\Phi_{0} \rangle,
\end{equation}
where $\Delta E_{\mu}$ is the energy change associated with the ionization process.
and ${\overline{H} = e^{-T} H e^{T}}-\langle \phi_{0}|e^{-T}He^{T}|\phi_{0}\rangle $ is a non-Hermitian operator.
This approach is usually known as EOM method for the excitation operators in analogy to the
Heisenberg's equation of motion. In the EOM-MBPT(2) and EOM-RPA methods, the matrix elements of the effective
Hamiltonian $\overline{H}$ are replaced accordingly in the above equation.

In the EOM-CCSD method, the cluster operators are defined as
\begin{eqnarray}
T &=& T_1 + T_2 = \sum_{i,a} t_{i}^{a} a_{a}^{+} a_{i} +\sum_{a<b} \sum_{i<j} t_{ij}^{ab} a_{a}^{+} a_{b}^{+} a_{i} a_{j} \ \ \ \ \ \ \\
&& \text{and} \nonumber \\
R_{\mu} &=& R_{1\mu} + R_{2\mu} = \sum_{i} r_{i} a_{i} + \sum_{i<j} \sum_{a} r_{ij}^{a} a_{a}^{+} a_{i} a_{j}, 
\end{eqnarray}
where $i,j$ indices are used for the occupied and $a,b$ are used for the virtual orbitals. 

The matrix elements of the effective Hamiltonian for the present ionization problem are constructed in the (1h-0p) 
and (2h-1p) space and diagonalized to get the desired roots. The Davidson algorithm \cite{50} has been implemented for the 
diagonalizing $\overline{H}$ . This is an iterative diagonalization scheme through which eigenvalues and eigenvectors are obtained.
It avoids computation, storage and diagonalization of the full matrix. The considered EOM-CC methods can be regarded as 
the diagonalization of the coupled cluster similarity transformed Hamiltonian in the CI configuration space. 

\section{Results and discussion}\label{sec4}

To test the performance of our newly implemented four component relativistic EOM-CCSD method,
we present numerical results of principal ionization potentials.
The calculations are performed for the closed-shell rare gas atoms ($\mathrm{He}$ through
$\mathrm{Rn}$), beryllium like ions (B through Ar and Kr), helium like ions (Li, Ne, Na, Ar, Kr)
along with $\mathrm{Na^{+}}$, $\mathrm{Al^{+}}$, $\mathrm{K^{+}}$, Be and Mg.
These calculations are compared with the results obtained using the
EOM-MBPT(2) and EOM-RPA methods to assess the role of electron correlation. All these results are compared with those of the measurements,
which are taken from the National Institute of Science and Technology (NIST)\cite{51} database.
For the construction of the single particle orbitals, we have used both even tempered (ET) and universal basis (UB) 
functions depending on the convergence of the results. For Be-like systems, we use UB basis with $\mathrm{\alpha_{0}}=0.004$ 
and $\mathrm{\beta}=2.23$. We have used ET basis for other atomic systems. The corresponding $\alpha_0$ and $\beta$ parameters
for ET basis for different atoms are given in Table \ref{tabgto}.
The use of total number
of orbitals generated at the SCF level is impractical in the
CC calculations as the contributions from the high lying 
orbitals are very small in the present calculations owing to their large energy values, we consider only the orbitals
that are significant of the calculations and they are termed as the active orbitals. In Table \ref{tabcorr}, we 
present (SCF) energy which is our zeroth energy (E$_{\text{DF}}^0$) and the correlation energies 
from the MBPT(2) (E$_{\text{corr}}^2$) and CCSD (E$_{\text{corr}}^{\text{CCSD}}$) methods along with the number of active
orbitals of different symmetries used in the calculations.

All the Gaussian type of functions generated at the SCF level are not important for the
ionization potential calculations. To investigate this, we have studied the convergence pattern of ionization potential
as a function of basis set through a series of calculations. Be atom is chosen for the convergence study.
We started our calculations with the 91 GTOs and gradually increased it to 118 GTOs.
It is found that the IP value of the 2s orbital changes in the order of $1\times 10^{-4}$ when the number of basis functions
increases from 91 to 118. The change is more for the 1s orbital and it is found to be $1.76\times 10^{-2}$, which is also 
in the accuracy limit of $0.01{\%}$. As we are more
interested in the valence ionization potential in the present work,
an active space of similar basis set is sufficient to construct the orthogonal space
for the inclusion of the correlation effects
for all the systems without compromising the desired accuracy . The results are given in Table \ref{tabbe}.

\begin{table}[t]
\caption{Ionization potentials (IPs) of noble gas atoms in eV using MBPT(2), RPA and CCSD methods in the EOM procedure.}
\begin{ruledtabular}
%\begin{tabular}{ l c c c c c }

\begin{tabular}{rrrrrr }

Atom  & Orbital & MBPT(2)  &  RPA &  CCSD  & NIST \cite{51}\\
\hline
$\mathrm{He}$    &\, $\mathrm{1s_{1/2}}$  &\,24.4560 &\,26.1086 &\,24.5802 &\, 24.5870     \\
                                                          \\
$\mathrm{Ne}$    &\,$\mathrm{2p_{3/2}}$   &\,21.4439 &\,25.5832 &\,21.4503 &\, 21.5642    \\
                 &\,$\mathrm{2p_{1/2}}$   &\,21.5499 &\,25.7096 &\,21.5560 &\,  21.6613                    \\
                 &\,$\mathrm{2s_{1/2}}$   &\,48.5478    &\,54.3474 &\,48.6207     &\, 48.4746                   \\
                 &\,$\mathrm{1s_{1/2}}$   &\,872.6377    &\,894.5355 &\,872.3581   &\,                     \\
                                                                           
                                                            \\              
                                                                           
$\mathrm{Ar}$    &\,$\mathrm{3p_{3/2}}$&\,15.8278  &\,18.0023   &\,  15.7951 &\, 15.7594 \\                                                         
                 &\,$\mathrm{3p_{1/2}}$&\, 16.0152 &\,18.2136   &\,  15.9817 &\, 15.9369           \\
                 &\,$\mathrm{3s_{1/2}}$&\,30.0706  &\,36.3317   &\,  30.0656  &\, 29.2390   \\
                 &\,$\mathrm{2p_{3/2}}$&\,250.1420  &\,261.8999    &\,249.7786 &\,        \\
                 &\,$\mathrm{2p_{1/2}}$&\,252.3757 &\,264.2143   &\, 252.0114  &\,              \\
%  \\
%                             \\
$\mathrm{Kr}$    &\,$\mathrm{4p_{3/2}}$  &\,14.1339 &\,15.8840   &\, 13.9963   &\,  13.9996     \\
                          \\
$\mathrm{Xe}$    &\,$\mathrm{5p_{3/2}}$  &\,12.3916 &\,13.7572   &\,12.1294    &\, 12.1298     \\
                    \\                       
$\mathrm{Rn}$    &\,$\mathrm{6p_{3/2}}$  &\,10.8604 &\,11.9900   &\, 10.5847   &\, 10.7485   \\
\end{tabular}
\end{ruledtabular}
\label{tabnob}
\end{table}
     
 We present the IP values of the helium like ions in Table \ref{tabhe}. All the results are sub-one percent
accurate and the result for $\mathrm{Li^{+}}$ is the most accurate and the least accurate is
0.15\% for $\mathrm{Kr^{32+}}$. This table shows that EOM-MBPT(2) results are always less than the EOM-CCSD
whereas the EOM-RPA method over estimates them compared to the NIST values. Also, the differences in the results 
between the EOM-MBPT(2) and EOM-CCSD are less than those of EOM-RPA and EOM-CCSD. The reason why the EOM-RPA 
calculations may be over estimating is that the 2h-1p block, which is the major source of non-dynamical correlations,
is not taken into account in this approach. The ground state wave function at the CCSD level is responsible for the 
major part of the dynamical correlations for which the EOM-MBPT(2) method seems to be a more valid approximation than
the EOM-RPA method. This suggests that the non-dynamical correlations are also important for the calculations of the
excited states. It is worth mentioning that the calculated EOM-CCSD IP results are larger than the NIST values
for $\mathrm{Ne^{8+}}$ onwards and the deviation gets larger as the nuclear charge increases.

\begin{table}[t]
\small
\caption{Ionization potentials (IPs) of $\mathrm{Na^{+}}$, $\mathrm{Al^{+}}$, $\mathrm{K^{+}}$, $\mathrm{Be}$ and $\mathrm{Mg}$ in eV using MBPT(2), RPA and CCSD methods in the EOM procedure.}
\begin{ruledtabular}
\begin{tabular}{rrrrrr}
Atom  & Orbital & MBPT(2) &  RPA &  CCSD & NIST \cite{51}\\
\hline
$\mathrm{Na^{+}}$    &\, $\mathrm{2p_{3/2}}$  &\,47.1177 &\,51.2511   &\,47.1286 &\,47.2863 \\
                     &\,$\mathrm{2p_{1/2}}$   &\,47.3000    &\,51.4556 &\,47.3105  &\, 47.4557        \\
                 &\,$\mathrm{2s_{1/2}}$       &\,79.9745    &\,85.4228  &\,80.0303  &\, 80.0741             \\   
                 &\,$\mathrm{1s_{1/2}}$       &\,1090.5239    &\,1112.3845&\,1090.3169&\,       \\

$\mathrm{Al^{+}}$    &\,$\mathrm{3s_{1/2}}$&\,18.6480  &\,19.1227 &\,18.8248&\,18.8285                 \\
                 &\,$\mathrm{2p_{3/2}}$&\,92.0692 &\,97.8533 &\, 91.9647& 91.7116 \\
                 &\,$\mathrm{2p_{1/2}}$&\,92.5141 &\,98.3291 &\, 92.4092&\, 92.1604 \\
                 &\,$\mathrm{2s_{1/2}}$&\,137.4759 &\,143.6488  &\, 137.4202&\,\\
                 &\,$\mathrm{1s_{1/2}}$&\,1582.3139 &\,1605.2003  &\,1582.0885&\,\\

$\mathrm{K^{+}}$    &\,$\mathrm{3p_{3/2}}$  &\,31.6687  &\,33.9023   &\, 31.6434&31.6249                  \\
                 &\,$\mathrm{3p_{1/2}}$     &\, 31.9497  &\, 34.2071  &\,31.9232&\,  31.8934                    \\
                 &\,$\mathrm{3s_{1/2}}$     &\,48.4814   &\,55.1066   &\,48.4795&\,  47.8182                     \\
                 &\,$\mathrm{2p_{3/2}}$     &\,309.0471   &\,320.6904   &\,308.7081&\,                   \\
                 &\,$\mathrm{2p_{1/2}}$     &\,311.9336   &\,323.6745   &\,311.5935&\,                        \\
                                                                                                                                                                                                              
$\mathrm{Be}$    &\,$\mathrm{2s_{1/2}}$&\,8.9442  &\,9.6603  &\, 9.3247&\, 9.3226 \\
                 &\,$\mathrm{1s_{1/2}}$&\,124.7175  &\,129.7139   &\,124.6463&\, 123.6344 \\
                 
$\mathrm{Mg}$    &\, $\mathrm{3s_{1/2}}$  &\,7.5057 &\,7.9519  &\, 7.6508 &\,7.6462 \\
                     &\,$\mathrm{2p_{3/2}}$&\,58.3976  &\,64.1697  &\, 58.2235    &  57.5603       \\
                 &\,$\mathrm{2p_{1/2}}$    &\,58.6898  &\,64.4875  &\, 58.5154    &\,  57.7983                    \\
                  &\,$\mathrm{2s_{1/2}}$   &\, 98.3383 &\,104.1001   &\,98.2824  &\, \\
\end{tabular}
\end{ruledtabular}
\label{tabot}
\end{table}

In Table \ref{belk}, we give the IP results for the beryllium like systems. The $2s$ valence IPs of these systems 
are in excellent agreement with the NIST values. Our results for the $1s$ orbital match reasonably well with the 
NIST data. 
We find that the relative average deviation of the IP values of beryllium like systems ($ \sim$ 0.03{\%}) are 
less than helium like ($0.05{\%}$) systems with reference to the the NIST data.

The results for the rare gas atoms are given in Table \ref{tabnob}. For the Kr, Xe, and Rn atoms, we have calculated
only the outer valence IPs. The most accurate EOM-CCSD result we obtain among them is for Xe atom. The $\mathrm{2p_{3/2}}$ 
valence ionization energy for the Ne atom differs from the experimental result by 0.1139 eV. The differences are 0.0357 
eV, 0.0033 eV and 0.0004 eV for the Ar, Kr and Xe atoms respectively for their valence orbitals. The reason for these
differences could be due to the possible double excitation character of the $p$ orbitals and it decreases 
along the group. The IPs of the EOM-CCSD method predominantly account for contributions from the single excitations and 
to some extent from the double excitations. The discrepancies could be mitigated on inclusion of the triple 
excitations in the ground as well as in the EOM part which is computationally very expensive for the relativistic 
calculations and are not incorporated in the present implementation. The deviation is 1.54{\%} for Rn atom,
which is expected as higher order relativistic effects are non negligible for heavy elements and also due to the finite
size of the basis sets.
 
In Table \ref{tabot}, we present the results for $\mathrm{Na^{+}}$, $\mathrm{Al^{+}}$, $\mathrm{K^{+}}$, Be and Mg. The 
largest deviation is found in the $\mathrm{2p_{3/2}}$ state of $\mathrm{Na^{+}}$ which is about to be 0.33{\%}. This
could be due to the possible dominance of the double excitations. In the case of  
$\mathrm{K{^+}}$  it is reduces to 0.05\% and for $\mathrm{Mg}$ it is 0.06\%.

\section{conclusion}\label{sec5}

The present work describes the four component relativistic implementation of the equation of motion coupled-cluster method
at the level of single and double excitations for the ionization problem in closed-shell atomic systems.  To test the 
reliability of this method, we have computed the ionization potentials of atomic systems from different groups
in the periodic table. The calculations are performed using EOM-MBPT(2) and EOM-RPA besides EOM-CCSD to understand 
the role of electron correlation at all the three levels of approximation. The second order many-body perturbation method is found to under
estimate the results, whereas the random phase approximation over estimates them. The EOM-CCSD results are in excellent agreement with the
NIST data wherever available. 
 
\section*{acknowledgements}
HP, NV and SP acknowledge the grant from CSIR XIIth five year plan project on Multi-scale Simulations of 
Material (MSM) and facilities of the Center of Excellence in Scientific Computing at CSIR-NCL. HP acknowledges the 
Council of Scientific and Industrial Research (CSIR) for his fellowship. SP acknowledges grant from DST, 
J. C. Bose fellowship project towards completion of the work. A part of the computations were carried out using
the 3T-Flop HPC cluster at Physical Research Laboratory.

\end{document}